\documentclass{aa}

\usepackage[varg]{txfonts}
\usepackage{graphicx}
\usepackage{array, booktabs, makecell}
\usepackage{siunitx}
\usepackage[version=4]{mhchem}
\usepackage{textcomp, gensymb}
\usepackage{amsmath}
\usepackage{amsfonts}
\usepackage{comment}
\usepackage{multicol}
\usepackage[utf8]{inputenc}
\usepackage[english]{babel}
\usepackage{natbib}
\bibpunct{(}{)}{;}{a}{}{,}
\usepackage[hidelinks,colorlinks=true,linkcolor=blue,citecolor=blue]{hyperref}
\usepackage[dvipsnames]{xcolor}

\begin{document}

\title{Shocks, clouds and atomic outflows in active galactic nuclei hosting relativistic jets} 

\author{Manel Perucho \inst{1,2}} 

\institute{
\inst{1} Departament d’Astronomia i Astrofísica, Universitat de València, C/ Dr. Moliner, 50, 46100, Burjassot, Val\`encia, Spain \\
\inst{2} Observatori Astronòmic, Universitat de València, C/ Catedràtic José Beltrán 2, 46980, Paterna, Val\`encia, Spain 
}

\date{Received xx / Accepted yy}

  \abstract
   {A number of observations have revealed atomic and/or molecular lines in active galaxies hosting jets and outflows. Line widths indicate outward motions of hundreds to few thousands of kilometers per second. They appear associated to the presence of radio emission in Gigahert-peaked spectrum (GPS) and/or compact steep spectrum (CSS) sources, with linear sizes $\leq 10\,{\rm kpc}$. Numerical simulations have shown that the bow shocks triggered by relativistic jets in their host galaxies drive ionisation and turbulence in the interstellar medium (ISM). However, the presence of atomic lines requires rapid recombination of ionised gas, which seems to be hard to explain from the physical conditions revealed so far by numerical simulations of powerful jets.}
{The aim of this paper is to provide a global frame to explain the presence of lines in terms of jet and shock evolution, and fix the parameter space in which the atomic and molecular outflows might occur.}
{This parameter space is inspired by numerical simulations and basic analytical models of jet evolution as a background.}
{Our results show that a plausible, general explanation involves momentum transfer and heating to the interstellar medium gas by jet triggered shocks within the inner kiloparsecs. The presence of post-shock atomic gas is possible in the case of shocks interacting with dense clouds that remain relatively stable after the shock passage.}
{According to our results, current numerical simulations cannot reproduce the physical conditions to explain the presence of atomic and molecular outflows in young radio-sources. However, I show that these outflows might occur in low-power jets at all scales, and predict a trend towards powerful jets showing lines at CSS scales, when clouds have cooled to recombination temperatures.}
   
\titlerunning{Shocks, clouds and atomic outflows in jetted AGN}
\authorrunning{Manel Perucho}
\maketitle

\section{Introduction}

Relativistic jets in AGN are produced in the environment of the central black holes. Our current understanding explains their formation in terms of magneto-hydrodynamical processes, via extraction of rotational energy either from a Kerr black-hole \citep{1977MNRAS.179..433B}, or from the inner accretion disk \citep{1982MNRAS.199..883B}. The jets are launched as sub-Alfv\'enic Poynting-flux dominated and accelerated by magnetic fields \citep{2004ApJ...605..656V,2007MNRAS.380...51K,2012rjag.book...81K} and internal energy \citep[][Ricci et al., submitted]{2007MNRAS.382..526P} to relativistic, superfast-magnetosonic speeds. It has been shown that the jets not only contribute to feedback at large scales and may participate in regulating star formation within their host galaxies \citep[see,e.g.][]{2009ApJ...699..525S,2019MNRAS.486.1509R}, but also may play a significant role in driving atomic and molecular outflows within the inner kiloparsecs from the AGN \citep[e.g.,][]{2018A&ARv..26....4M,2020A&A...644A..54S,2021AN....342.1135M,2021AN....342.1200T,2023A&A...674A.198K}.

VLBI observations of compact, young radio sources revealed advance velocities of $\sim 0.2\,c$ \citep[e.g.,][]{1998A&A...337...69O,2003PASA...20...69P}. Such velocities probably correspond to powerful jets. Numerical simulations of jet propagation within the inner kiloparsec of active galaxies have shown that they do expand at supersonic velocities and trigger strong shocks in the interstellar medium \citep[see, e.g.,][]{2011ApJ...728...29W,2012ApJ...757..136W,2016MNRAS.461..967M,2018MNRAS.475.3493B,2018MNRAS.479.5544M,2021AN....342.1171P,2022MNRAS.511.1622M}. Simulations also show that lateral expansion is slower by a factor of 3-5, as derived from aspect ratios, so the radial expansion velocity can be $\simeq 0.04 - 0.06\,c$. Such strong shocks increase the temperature of the gas to values high enough to completely ionise hydrogen atoms. Post-shock temperatures of $10^8 - 10^{10}$~K can be easily achieved in this region, as we also justify in this work. In contrast, simulations of low power jets ($L_j \simeq 10^{42}\,{\rm erg/s}$) present advance velocities $\leq 10^{-3}\,c$, and temperature jumps $\leq 2$ \citep{2014MNRAS.441.1488P} to values $\sim 10^7$~K (in the hot ISM component).

However, radio observations of atomic lines show the presence of HI outflows in compact radio sources, i.e., Gigahertz-peaked spectrum (GPS) sources \citep{Holt11}, with linear sizes $\leq 1\,{\rm kpc}$, and compact steep spectrum (CSS) sources, $\leq 10\,{\rm kpc}$ in size \citep[see, e.g.][and \cite{2018A&ARv..26....4M} for a review]{ODea02,labiano05,Shih13,reynaldi16,2016A&A...593A..30M,2019MNRAS.489.4944Z,2021A&A...647A..63S}. The presence of HI requires gas temperatures $\leq 1.5\times10^5\,{\rm K}$ in the case of collisional equilibrium, or even smaller in the case of thermodynamical equilibrium, i.e., at least four orders of magnitude below the shocked gas temperature in powerful jets \citep[e.g.,][]{2021AN....342.1171P}. The inferred velocities range from several hundreds to few thousands of km/s. In \citet{Holt11}, the authors point out that the line widths reveal larger velocities in two observed GPS galaxies than in CSSs, and that they are compatible with an in situ shock-cloud collision, which has also been reported by \citet{2020MNRAS.497.5103C} for detected OIII lines. However, this is not obvious in all sources.   

Although the aforementioned numerical simulations have tackled the problem of jet evolution inside the host galaxy, the authors have been forced to focus either on the inner kiloparsec (GPS region) or beyond (CSS region) by computational limitations, so the transition has barely been discussed in terms of heating and cooling of the ISM gas. Furthermore, none of the simulations had the required resolution to study the evolution of shocked clouds (see, however, Mandal et al., in preparation).

\citet{1997ApJ...485..112B,2003PASA...20..102B} presented a model based on \citet{1996cyga.book..209B} that successfully explained a number of properties of GPS sources. The authors show that radiative shocks can create an ionised shocked ISM that free-free absorbs low frequency radiation and explain the evolution of the peak frequency with size in these young radio-galaxies \citep{1990A&A...231..333F,1997AJ....113..148O,1998A&AS..131..303S}. Those theoretical papers focus on high power jets ($10^{45-46}\,{\rm erg/s}$). 

This paper takes a simple analytical approach to tackle the discrepancy between physical conditions derived from numerical simulations and those required for the presence of HI outflows in host galaxies \citep[see, e.g.,][and references therein for a detailed study of the physics of two-fluid shocked plasmas]{2021A&A...645A..81S,2021MNRAS.506.1334S}. The model presented here is motivated by results from numerical simulations of relativistic outflows \citep[see, e.g.][for other analytical models also driven by simulations and \citet{2023Galax..11...87T} for a review]{2018MNRAS.475.2768H,2023MNRAS.518..945T} and focuses on the role of shocks on the environment, extending the study to low power radio-galaxies. The shocks are shown to be non radiative in the case of powerful jets, but to become radiative at a few kpc for low power jets. The model shows that atomic outflows are compatible with conditions produced by low power jets at all scales, but require further propagation time and cloud stability to allow for gas cooling in the case of powerful sources.

These results would point towards more powerful radio-sources showing atomic lines when they have propagated longer distances, but low power radio-sources presenting them at all scales, including smaller linear sizes. Altogether, this conclusion sets a plausible general scenario to explain observations of atomic outflows at all scales, depending on the jet power, either within or beyond the inner kiloparsec.

The paper is structured as follows. In section~2, I present the physical arguments that could explain the presence of atomic lines in active galaxies based on shock-heating and estimated cooling times. In section~3, I discuss the implications for both low and high power radio-galaxies, the caveats of the simple analytic approach and the implications for our understanding of jet physics and feedback at larger scales. Finally, in section~4 I summarise the main conclusions of this paper.

\section{Shocks: heating and cooling} \label{sec:heating}
\subsection{The heating} \label{sec:model}

We focus here on the lateral expansion of jet-triggered shocks (i.e., misaligned with the jet direction), which affects a large volume of the ISM. 
The shock propagates into the two-phase ISM, with a hotter, dilute medium with temperatures $\sim 10^6-10^7\,{\rm K}$ and number densities $\sim 0.1-1\,{\rm cm^{-3}}$, and a colder, denser component in clouds, with temperatures $\sim 10-1000\,{\rm K}$ and densities $\sim 10^2-10^3\,{\rm cm^{-3}}$. The shocked ISM medium surrounds the shocked jet gas (or cocoon), with very high temperatures and lower densities \citep[see, e.g.,][]{2011ApJ...728...29W,2016MNRAS.461..967M,2018MNRAS.479.5544M,2021AN....342.1171P}. Nevertheless, our interest is focused on the shocked ISM-ISM interaction region \citep[see][for the case of a cloud entering and interacting with the jet cocoon]{2007MNRAS.376..465K,2008MmSAI..79.1162K}.  

 
The following expressions (jump conditions in the shock reference frame) give the post-shock pressures and temperatures:
\begin{equation}\label{eq:mass}
\rho_1\,v_1\,=\,\rho_2\,v_2,
\end{equation}
\begin{equation}\label{eq:momentum}
v_2 \,=\,\frac{P_1\,-\,P_2}{\rho_1\,v_1} \,+\,v_1,
\end{equation}
\begin{equation} \label{eq:energy}
v_1^2 \,=\,\frac{1}{2\,\rho_1}\left[ (\Gamma+1)\,P_2+(\Gamma-1)\, P_1 \right]
\end{equation}
with subscripts $1$ and $2$ standing for pre and post-shock states of the interstellar medium (ISM), respectively, and where $P$ represents pressure, $\rho$ the gas rest-mass density, $v$ the velocity in the shock propagation direction, and $\Gamma$ the adiabatic exponent of an ideal gas. 

In the reference frame of the shocked ISM, $v_1 = v_s$, the shock velocity. For an assumed ideal gas, we also have $P/\rho\,=\, k\,T/\mu m_H$, where $k$ is the Boltzmann constant, $\mu$ is the mean molecular weight, and $m_H$ is the atomic hydrogen mass. The previous expressions allow us to derive the shock advance velocity by imposing the ambient density and both pre and post-shock pressures.

\subsubsection{The hot phase}

In the case of AGN bow-shocks, the post-shock pressure is rapidly homogenised in the shocked hot phase because of the large sound speed achieved by the gas, as it has been shown by numerical simulations. \citet{2017MNRAS.471L.120P} also showed that the following expression gives an order-of-magnitude estimate of post-shock pressure in relativistic jets:

\begin{equation} \label{eq:bp}
P_2\,\leq\,\frac{L_j\,t}{V}, 
\end{equation}
where $L_j$ is the jet power, and $V$ is the shocked volume.\footnote{The expression implicitly assumes that all injected power is converted into internal energy driving the lobe expansion, which is valid for relativistic (hot and fast) outflows. Thus, this represents an upper limit for pressure, which propagates as an upper limit for the derived post-shock temperatures.} The ISM pressure can be taken to be the usually estimated value $P_1\,\simeq\,10^{-10}\,{\rm dyn\,cm^{-2}}$. Therefore, if we are able to estimate the shocked volume with time, we can get an approximate value for the pressure jump. We can do this in a very raw way, taking the shocked volume to be a cylinder, with radius $r_s$ and length $l_s$ \citep[a usual approximation used in analytical models, e.g.,][]{1989ApJ...345L..21B}: $V\,\sim\, \pi\, r_s^2 \, l_s$. Considering the mean head advance velocity at small evolution times (within the host galaxy), we can write $r_s \,\sim \, v_s\, t$ and $l_s \,\sim \, v_h\, t$, where $v_s$ is the lateral shock velocity, which is the one we are interested in, and $v_h$ is the head advance velocity. As stated above, $v_s$ is smaller than $v_h$. Self-similarity appears as a result of keeping the ratio $v_h/v_s$ constant. \citet{1997MNRAS.286..215K} showed that this is a consequence of adiabatic expansion of jet-inflated cocoons. This is supported by observed lobe/hot-spot$-$distance evolution throughout the inner kpc \citep{2002ApJ...568..639P,2003PASA...20...69P,2008ApJ...687..141K} in the case of Compact Symmetric Objects. With this information, we can write the expression for the volume as a function of time:   
\begin{equation}\label{eq:vol}
V(t)\,\sim\,\pi\,v_{0,s}^2\,v_{0,h}\,t^{3-3\alpha}.
\end{equation}   


For the first time step we can safely assume strong shock conditions $P_2\,=\,\rho(r)\,v_s^2$ (where $\rho(r)$ is the ambient density). With this, we obtain
\begin{equation}
v_{s,0}\,=\,\left(\frac{L_j}{3\,\pi \rho_0 t_0^2} \,\right)^{1/5}. 
\end{equation}

\begin{figure*}[t]
    \centering
    \includegraphics[width=\columnwidth]{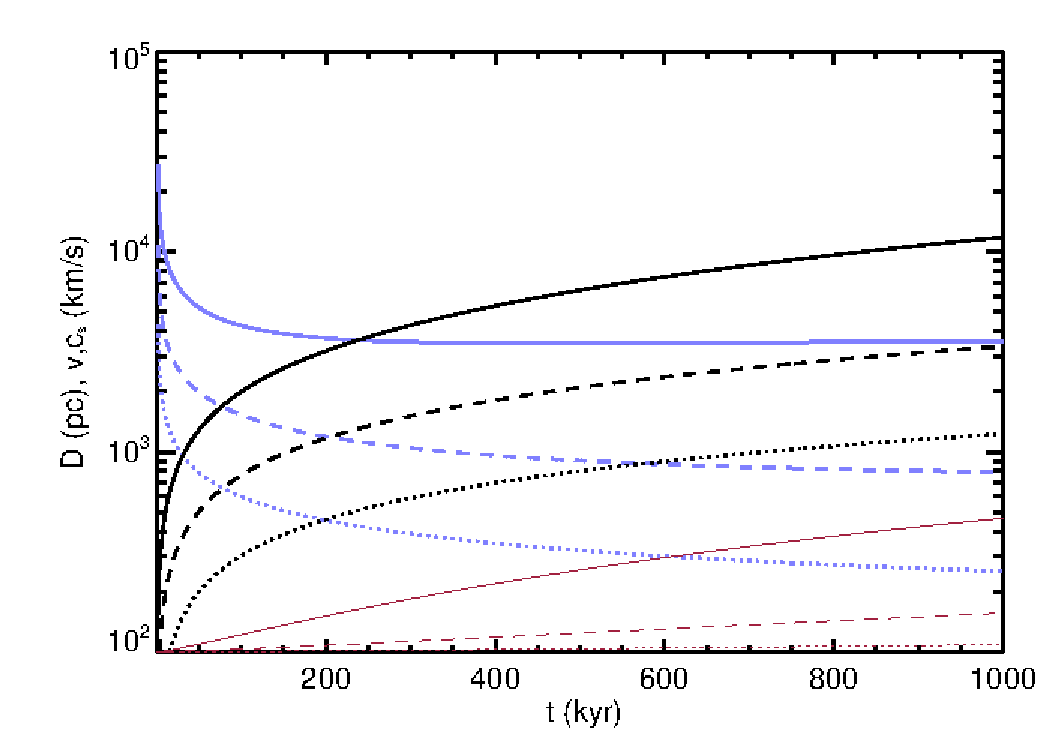}\, \includegraphics[width=\columnwidth]{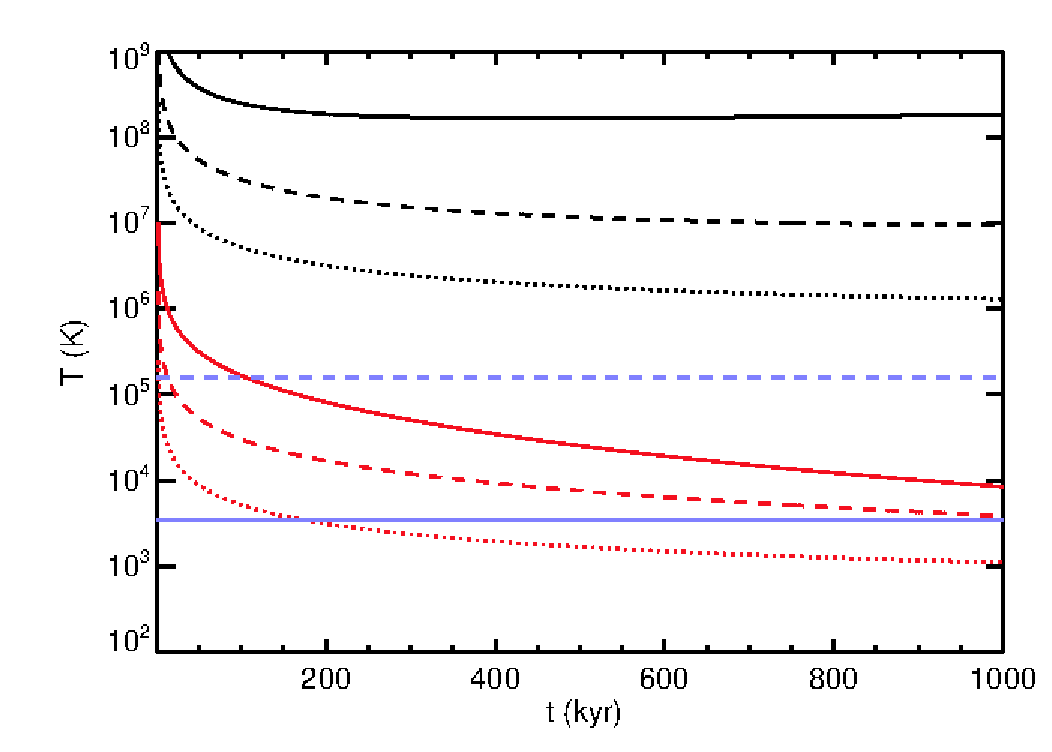}
    \caption{Left panel: radial shock velocity ($v_s$, violet) and axial position (black) for jet powers $10^{46}$ (solid lines), $10^{44}$ (dashed), and $10^{42}\,{\rm erg/s}$ (dotted). The thin, dark red lines indicate the ISM sound speed at the axial position. Right panel: post-shock temperatures for same jet powers as the left panel in the case of a medium density following Eq.~\ref{eq:rhoa} (black lines), i.e., a hot, diffuse ISM, and a medium with constant number density $10^3\,{\rm cm^{-3}}$, i.e., cold clouds. Horizontal lines indicate hydrogen ionization temperatures in local thermodynamical equilibrium, $3500\,$K (solid), and collisional equilibrium, $1.58\times10^5\,$K (dashed). In these clouds, the sound speed is $3\,{\rm km/s}$. The temperatures have been derived for ionised hydrogen in Eq.~\ref{eq:temp}.}
    \label{fig:lines1}
\end{figure*}


Finally, we assume a typical ambient medium density profile 
\begin{equation}\label{eq:rhoa}
\rho(r)\, =\, \rho_0 \, \frac{1}{1\,+\,\left(\frac{r}{r_c}\right)^2},
\end{equation}
and $\rho_0\,=\,1.67\times10^{-24}\,\rm{g/cm^3}$.

We can provide a general expression to estimate post-shock gas temperature by starting from Eq.~\ref{eq:bp}. Taking an initial volume of $\sim 100\,{\rm pc}^3$ as a reference at $t\sim 10^3\,$yr, we get:

\begin{equation} \label{eq:bp0}
P_2 (t_0)\,\simeq \, 3\times10^{-7}\, \left(\frac{L_j}{10^{45}\,{\rm erg/s}}\right)\, \left(\frac{t}{10^3\,{\rm yr}}\right)\, \left(\frac{V}{(100\,{\rm pc})^3}\right)^{-1}\,{\rm dyn\,cm^{-2}},
\end{equation}
i.e., a factor $10^3$ over the ISM pressure (validating the strong shock hypothesis at $t_0$). The evolution of pressure with time can then be written from the previous expression and the time-dependency obtained for the volume (see Eq.~\ref{eq:vol}):


\begin{equation} \label{eq:bpt}
P_2 (t)\,\simeq\, 3\times10^{-7}\, \left(\frac{L_j}{10^{45}\,{\rm erg/s}}\right)\, \left(\frac{t}{10^3\,{\rm yr}}\right)^{-2+3\alpha}\, {\rm dyn\,cm^{-2}},
\end{equation}
where the constants in Eq.~\ref{eq:vol} have been absorbed in the initial volume taken to estimate $P_2 (t_0)$. 

Post-shock temperature can be estimated from (for a strong, non-relativistic shock):
\begin{equation} \label{eq:temp}
T_2 (t)\,=\,\frac{P_2(t)}{\rho_2}\,\frac{\mu\,m_H}{k}\,\simeq \frac{P_2(t)}{4\,\rho_1}\,\frac{\mu\,m_H}{k}.
\end{equation} 
Taking $\rho_1 \, =\,\rho_0  \simeq 1 \,m_p \,{\rm cm^{-3}}$ as a reference value for the unperturbed ISM, we obtain: 
\begin{equation} \label{eq:tempt}
T_2 (t)\,\simeq \, 2\times10^9 \left(\frac{L_j}{10^{45}\,{\rm erg/s}}\right) \left(\frac{\rho_1}{1\,m_p\,{\rm cm^{-3}}}\right)^{-1}\, \left(\frac{t}{10^3\,{\rm yr}}\right)^{-2+3\alpha}\, \rm{K}.
\end{equation} 

The equation shows that the post-shock temperature is linearly dependent on the jet power and inversely proportional to ISM gas density. The dependence on time is controlled by exponent $\alpha$ for velocity evolution. Plausible values range from $\alpha=0$, which is a consequence of a very steep density profile ($\rho(r)\propto r^{-2}$) or a buoyantly propagating bubble with constant velocity, and $\alpha \leq 1$ for low-power and or mass-loaded jets \citep{2011ApJ...743...42P,2014MNRAS.441.1488P}. Self-similarity implies $\alpha \sim 0.4$ \citep{1974MNRAS.166..513S,1997MNRAS.286..215K}, so post-shock temperature would fall with time as $t^{-0.8}$. The sensitivity of $T_2$ to source age would be damped for $\alpha \simeq 0.5-0.6$. On the contrary, it would be stronger if velocity is constant ($\alpha=0$ would make $T_2 \propto t^{-2}$, owing to a fast pressure decrease) or if it fell rapidly ($\alpha \simeq 1$ would make post-shock pressure become proportional to $t$, due to the increase in pressure if the advance is slowed down). For moderate values of $\alpha$ as that given by self-similarity, post-shock temperature might fall approximately as $t^{-1}$ when propagating through an ambient medium with constant density, i.e., the galactic core. 

Therefore, beyond this initial set of values, we could impose self-similar expansion and strong-shock conditions all along the studied evolution. In this case $\alpha = 2/5$ and the evolution is determined by the previous equations. However, as we will see, the latter assumption could be not fulfilled for low power jets. In order to avoid this problem, we can proceed in a different way: the shock advance velocity in the radial direction at time $t_i$, $v_s(t_i)$, can be estimated from Eq.~\ref{eq:energy}, i.e., skipping the strong-shock assumption. Once $v_s(t_i)$ is derived (for time $t_i$), we can obtain $V(t_{i+1})$ assuming constant expansion velocity for a sufficiently small time-step.

However, Eq.~\ref{eq:vol} requires not only $v_s$ but also $v_h$ to be known. We have seen that self-similarity can be a good assumption through the inner region and, in addition, the lobe length-to-radius ratio is $\leq 5$, so we can assume a constant relation between both expansion velocities, e.g., $v_h \simeq 3 v_s$. Although this assumption is probably weaker later on, when the ratio can decrease due to jet head deceleration, $v_h/v_s$ must be always larger than 1, so we can impose the same velocity ratio along the whole studied distance range. Detailed analytical modelling by \citet{2018MNRAS.474.3361T} showed that, in realistic environments, the ratio of jet head to transverse expansion velocity only increases moderately (factor ~1.5 on scales of 100s of kpc; see Turner et al.’s Figures 8 and 9). As shown in that paper, these analytical results are consistent with numerical simulations, and also observations of lobed 3C radio sources. Furthermore, this estimate cannot introduce significantly larger errors than, for instance, considering a cylindrical volume for the shocked region.

Finally, once we obtain the value of $V(t_{i+1})$, we find $P_2(t_{i+1})$ from Eq.~\ref{eq:bp}, and proceed to compute $v_s(t_{i+1})$. In this way, we can avoid the dependency on $\alpha$ for the expansion velocity.

Figure~\ref{fig:lines1} shows the evolution of shock position and velocity along the jet propagation direction (left panel) and post-shock temperatures (right panel) for different jet powers. The right panel also displays post-shock temperatures for media with constant number density $10^3\,{\rm cm^{-3}}$ (and pressure $P_1$) in order to show the values attained inside clouds shocked at different locations within the host galaxy (see Sect.~\ref{sec:cold}). We have set $t_0=10^3$~yr and have used constant ambient pressure $P_1=10^{-10}\,{\rm dyn/cm^2}$. 

\begin{figure}[t]
    \centering
    \includegraphics[width=\columnwidth]{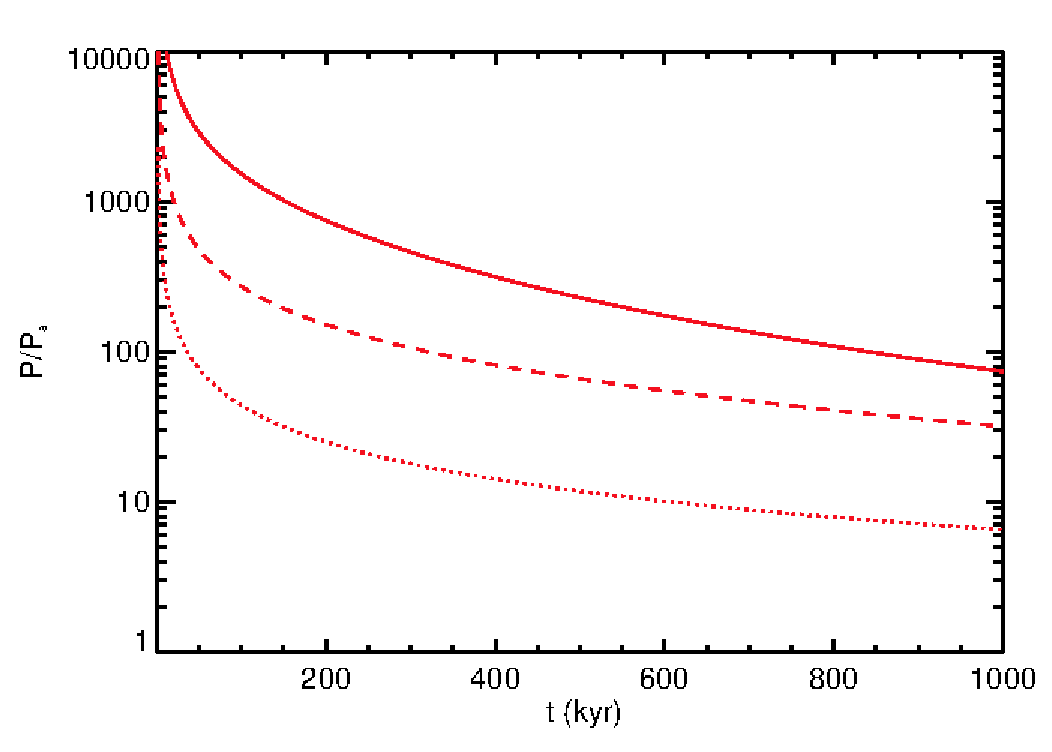}
    \caption{Lobe overpressure for jet powers $10^{46}$ (solid lines), $10^{44}$ (dashed), and $10^{42}\,{\rm erg/s}$. {\bf The black lines for the hot, diffuse ISM, and red lines correspond to cold, dense clouds with number density $10^3\,{\rm cm^{-3}}$ are overlapped.}}
    \label{fig:linesp}
\end{figure}

The right panel in Fig.~\ref{fig:lines1} shows that post-shock temperatures reached by the hot, dilute ISM agree with those obtained for powerful radio-galaxies from numerical simulations. The ratios $P_2/P_1$ for pressure jumps across the shocks (see Fig.~\ref{fig:linesp}) indicate that overpressure goes below a factor 10 only in the case of the low-power radio galaxies after $500\,{\rm kyr}$, when the jet head is at $\simeq 1\,{\rm kpc}$, and reach values $\simeq 7$ at $1{\rm Myr}$ ($D\simeq 1.2\,{\rm kpc}$). Therefore, the strong shock approximation would remain valid throughout most of the shown profiles for powerful jets, but low power radio galaxies may transit to weaker (and radiative, see Sect.~\ref{sec:cool}) shocks within the inner 2~kpc of propagation.

\subsubsection{The cold phase} \label{sec:cold}

The red lines in the right panel of Fig.~\ref{fig:lines1} and Fig.~\ref{fig:linesp} show the estimated values of temperature and overpressure for shock-cloud interactions, assuming $n\,=\,10^3\,{\rm cm^{-3}}$, for the three considered jet powers. 

We have assumed $v_{1,c} \sim (\rho_1/\rho_{1,c})^{1/2} v_1$ (subscript $c$ stands for clouds, and no subscript for the values derived for the hot phase in the previous section). Taking $P_{1,c}\,=\, P_1$ and $\Gamma=5/3$ for a cold, monoatomic gas (hydrogen), we can derive $v_{2,c}$ by rearranging the jump Eqs.~\ref{eq:mass}-\ref{eq:energy}, and then $P_{2,c}$.



\begin{figure}[t]
    \centering
    \includegraphics[width=\columnwidth]{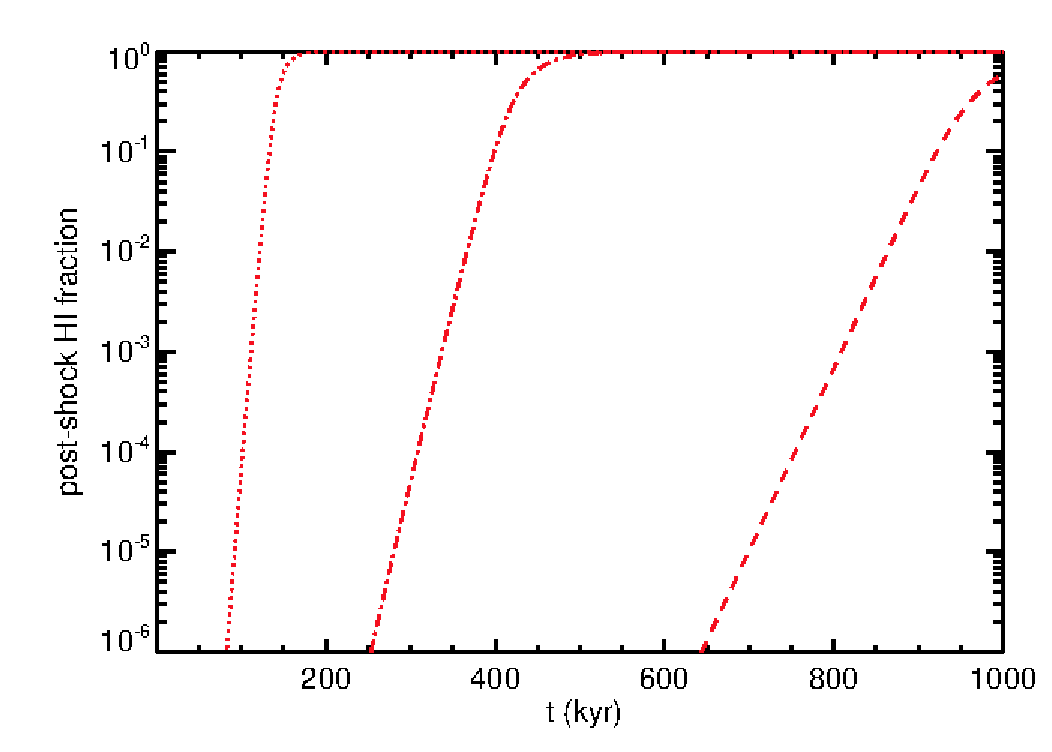}
    \caption{HI fraction in shocked clouds in LTE for jets with powers $10^{44}$ (dashed), and $10^{42}\,{\rm erg/s}$ (dotted). The intermediate case of $10^{43}\,{\rm erg/s}$ (dash-dot line) has been also included in this plot.}
    \label{fig:saha}
\end{figure}

Assuming collisional ionization, which is expected to be a valid approximation for diffuse, astrophysical plasmas \citep{2003adu..book.....D}, hydrogen ionising temperatures are $\sim 10^5\,$K. Figure~\ref{fig:lines1} shows that, for $L_j\,=\,10^{46}\,{\rm erg/s}$ temperatures $\simeq 10^5\,$K are reached in shocked clouds up to $\simeq 100\,{\rm kyr}$ ($D\simeq 2\,{\rm kpc}$). In the case of $L_j\,=\,10^{44}\,{\rm erg/s}$, this happens before $50\,{\rm kyr}$ ($D\simeq 300\,{\rm pc}$). Finally, low power jets may barely reach ionising post-shock temperatures in dense clouds within the inner tens of parsecs.

Although local thermodynamical equilibrium (LTE), in which the ionised states follow Boltzmann's law \citep{2003adu..book.....D}, is probably not fulfilled in diffuse astrophysical plasmas, we can explore this possibility, too, for the case of clouds. Using the post-shock values derived for clouds and Saha equation, we can derive the degree of ionization in shocked clouds. In this case, ionising temperatures are $\simeq 3000\,$K. The results are shown in Fig.~\ref{fig:saha} for $L_j\,=\,10^{42}$, $10^{43}$ and $10^{44}\,{\rm erg/s}$. Low power jets would be unable to ionise hydrogen in cold, dense clouds, beyond a few hundred parsecs, whereas the first traces of HI in shocked clouds would appear after $D\simeq 3\,{\rm kpc}$ in $10^{44}\,{\rm erg/s}$ jets. Post-shock temperatures are incompatible with the presence of non-ionised HI in shocked clouds in LTE for higher powers. Therefore, higher power jets would require efficient cooling to explain the presence of atomic outflows within the inner kiloparsecs of propagation, if LTE is assumed.

These values are only a basic approximation because shocks in clouds are probably radiative (see Sect.~\ref{sec:cool}), allowing for fast cooling. As an opposing, heating mechanism, the development of instabilities and mixing with the hotter environment could change this picture, as we later discuss. 

\subsection{The cooling} \label{sec:cool}

\begin{figure*}[t]
    \centering
    \includegraphics[width=\columnwidth]{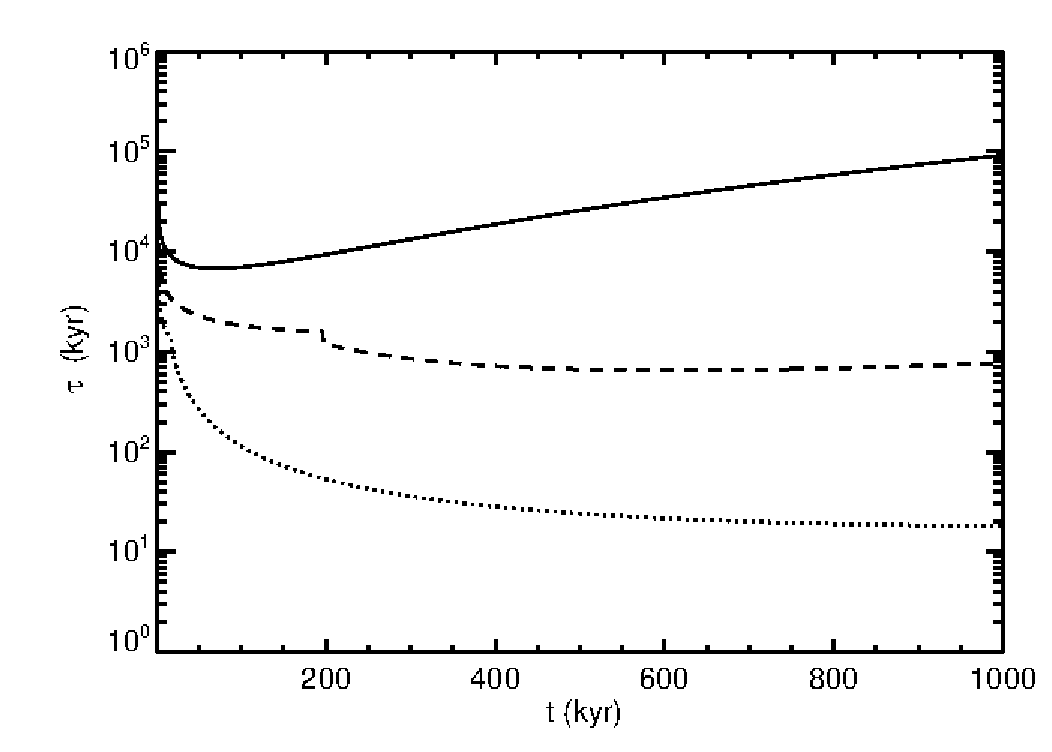}\, \includegraphics[width=\columnwidth]{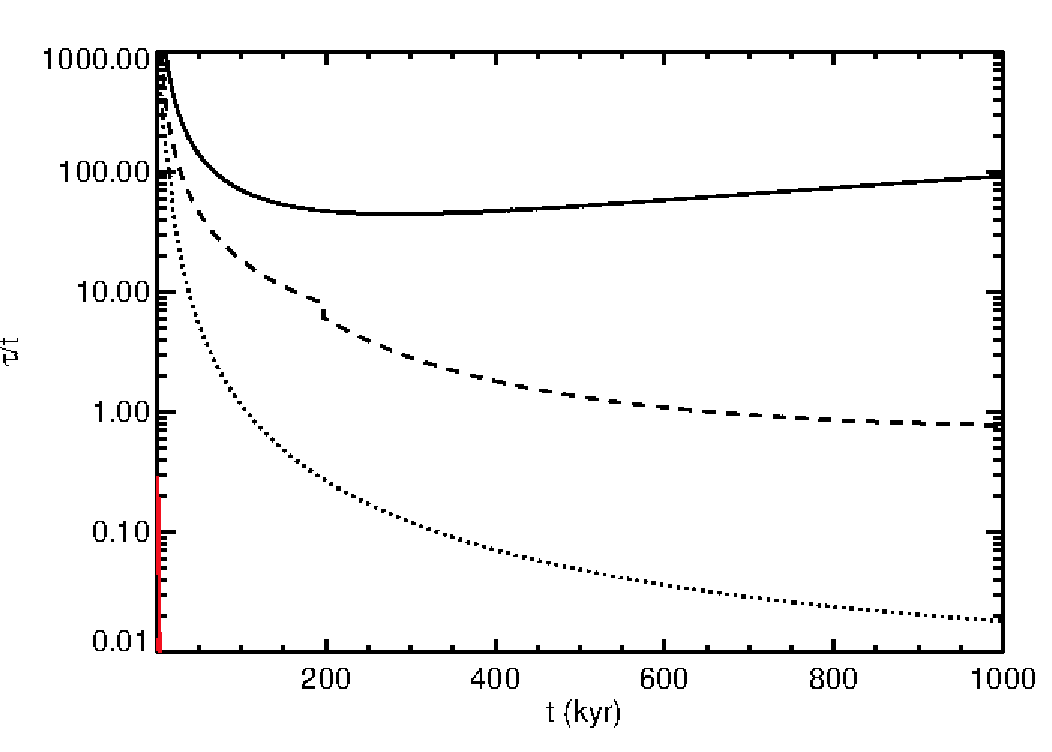}
    \caption{Cooling times for the same cases given in Fig.~\ref{fig:lines1} (left panel) and relative values with respect to the evolution times (right panel). Solid lines represent the powerful, $10^{46}\,{\rm erg/s}$ case, dashed lines the $10^{44}\,{\rm erg/s}$ case, and dotted lines the $10^{42}\,{\rm erg/s}$. Calculations are shown for the shocked hot, dilute ISM described by Eq.~\ref{eq:rhoa}.} 
    \label{fig:lines2}
\end{figure*}

As we have stated in the Introduction, outflows of atomic gas have been observed both in GPS ($l_s \leq 1\,{\rm kpc}$) and CSS sources ($l_s \leq 10\,{\rm kpc}$). The values obtained for temperature in the previous section (see Eq.~\ref{eq:tempt}) and in numerical simulations of powerful jets certainly prohibit the presence of atomic hydrogen when the radio-galaxy extends over the inner hundreds of parsecs. Efficient cooling must take place to explain the presence of atomic outflows correlated with the presence of more powerful ($L_j \geq 10^{43}\,{\rm erg/s}$) relativistic jets. 

Dense clouds are expected to be located within the inner kiloparsec of the host galaxy. In the case of powerful jets that ionise most of the shocked ISM, we then expect the observed atomic outflows to be formed by the recombination of those ionised atoms, once the radio-source has evolved to larger scales. 

According to cooling theory \citep[see, e.g.,][]{1998MNRAS.298.1021M,2003adu..book.....D,2011piim.book.....D} the cooling function, $\Lambda/n_H n_e$ can be approximated to $1.1\times10^{-22}\,(T/10^6)^{-0.7}\,{\rm erg\,cm^{3}\,s^{-1}}$ for gas temperatures between $10^5\,$K and $2\times10^7\,$K, beyond which it is dominated by Bremsstrahlung ($2.3\times10^{-24}\,(T/10^6)^{1/2}{\rm erg\,cm^{3}\,s^{-1}}$), for number density $n_H = 1\,{\rm cm^{-3}}$, and, at lower temperatures (between $10^4\,$ and $10^5\,$K), it becomes $7\times10^{-27}\,T{\rm erg\,cm^{3}\,s^{-1}}$. Below $10^4\,$K, we use the cooling function for recombination of hydrogen atoms \citep[$2.39\times10^{-27}\,T^{1/2}{\rm erg\,cm^{3}\,s^{-1}}$, see][]{2015A&A...580A.110V}. Using the values for post-shock densities and temperatures obtained in Sect.~\ref{sec:model}, we can compute the cooling times of the shocked gas. 

The cooling times are computed as follows: 
$$\tau\,=\, \frac{(\Gamma - 1)\,P_2}{n_2^2\, 2.3\times10^{-24}(T_2/10^6)^{0.5}}\,{\rm s}, \quad T_2 > 2\times10^7{\rm K},$$
$$\tau\,=\, \frac{(\Gamma - 1)\,P_2}{n_2^2\, 1.1\times10^{-22}(T_2/10^6)^{-0.7}}\,{\rm s}, \quad 10^5\,{\rm K} < T_2 <2\times10^7{\rm K},$$ 
$$\tau\,=\, \frac{(\Gamma - 1)\,P_2}{n_2^2\, 7.0\times10^{-27} T_2}\,{\rm s}\, \quad 10^4\,{\rm K} < T_2 <10^5{\rm K},$$
$$\tau\,=\, \frac{(\Gamma - 1)\,P_2}{n_2^2\, 2.39\times10^{-27} T_2^{0.5}}\,{\rm s}\quad T_2 < 10^4{\rm K}.$$

The results are shown in Fig.~\ref{fig:lines2}, where we observe that low-power jets present cooling times for the shocked hot, dilute ISM of the order or less than evolution times after $10^5\,$yr, when they become radiative. The cooling times derived for cold, dense clouds are shorter than 1~kyr and are thus not shown. This means that cooling in the clouds is basically instantaneous in relative terms. Therefore, although fast jet expansion and high post-shock temperatures would make it difficult to find atomic hydrogen within the inner kiloparsec in powerful jets, such short cooling times could make recombination plausible in shocked clouds even before a linear size of 2~kpc is reached.



\section{Discussion} \label{sec:discussion}

\subsection{Low power radio galaxies}

We have seen that the bow shock can weaken and cease to be 'strong' beyond a few kpc for jet powers $L_j\,<\, 10^{43}\,{\rm erg/s}$  (see Figs.~\ref{fig:lines1} and \ref{fig:linesp}). Even though post-shock temperatures are high for the hot, dilute ISM phase, cooling times become similar to source age when the tip of the bow shock is still located within the inner kpc, and keep decreasing farther on. In the case of cold, dense clouds, however, the post-shock temperatures are always $\leq 10^5$~K. This means that 1) it is possible that part of the atomic gas is not ionised by the shock, and 2) cooling times are basically instantaneous compared to the evolution time-scales. 

As the shock weakens, post-shock temperature and velocity become smaller for the hot, dilute gas surrounding the clouds. Because the difference in conditions inside and outside shocked clouds would be then alleviated, shocked clouds could be more stable within this environment than in that generated by stronger shocks. In addition, even if mixing occurred, the lower surrounding temperatures would still allow for rapid cooling. This is, however, to be studied by devoted numerical simulations.

Nevertheless, the simple analytical approach that we have used already indicates that the presence of atomic gas with outward motions in shocked regions within radio galaxies could be relatively easily explained.

\subsection{Powerful radio galaxies}

In Sect.~\ref{sec:cool} we have seen that, despite the high temperatures achieved by the ISM after the passage of shocks, large density regions could allow fast enough cooling and, therefore, the presence of atomic (and molecular) gas within the inner kiloparsec(s) from the active nucleus. In addition, this gas has an outward momentum that can be aligned with the jet propagation direction or be perpendicular to it because the bow shocks propagate in all directions, in accordance with observations. The observed lines reveal radial velocities of hundreds up to few thousand km/s, and these velocity fields are indeed reproduced in numerical simulations \citep{2021AN....342.1171P,2022MNRAS.511.1622M,2022NatAs...6..488M}.

The higher temperatures achieved by the shocked gas in powerful radio-sources and thus longer cooling times could lead us to expect a bias in the detection of atomic outflows towards larger distances to the nucleus in this case. Furthermore, (shocked) cloud stability has to be taken into account: disruption and mixing with the hotter environment would significantly delay cooling times/distance. An \emph{a fortiori} argument favouring cloud stability conditions is given by observations. Clouds must retain a certain degree of coherence after shock passage if atomic and molecular lines are observed within shock affected regions in radio-galaxies. Otherwise, this would be difficult to explain for relatively powerful radio-galaxies. 

In a recent paper, \citet{2021A&A...645A..81S} explore the physics of shock propagation through partially ionised plasmas and show that post-shock equilibrium in such systems can lead to rapid cooling and lower post-shock than pre-shock temperatures. If this scenario is confirmed and generalised in the context of AGN shocks and clouds, it could provide an easy explanation to the presence of atomic lines associated with young radio-galaxies.

Unstable processes evolve on timescales of $\sim R/c_s$, where $R$ is the cloud radius and $c_s$ its sound speed. The relatively low temperatures kept by cloud material (both unshocked and shocked) ensures relatively low sound speeds. 
For a cloud size of 10~pc and a sound speed (after shock passage) of $10^2-10^3\,{\rm km/s}$ dynamical times become $\sim 10^4-10^5$~yr, i.e. of the order of typical evolution times shown in Fig.~\ref{fig:lines1} and certainly longer than cooling times. There could thus be time for cooling and recombination prior to eventual disruptive processes. So, even if mixing would induce further ionization, there could be a long enough time interval between shock passage and disruption when atomic lines could be observed. Another physical mechanism to be taken into account is enhanced clumping induced by self-gravity could also contribute to further cooling and stability (Mandal et al., in preparation).

As stated above, the development of instabilities would favour mixing with the hotter, shocked surrounding gas, acting against cooling and recombination in the region. Numerical simulations of radio source evolution \citep[see, e.g.][]{2016MNRAS.461..967M,2021AN....342.1171P} do not show any hints of post-shock gas colder than $10^8\,$K, although the densest clouds in set-ups do have densities $\geq 100\,{\rm cm^{-3}}$. The reason is numerical: the clouds in such simulations do not involve many cells ($\leq$ 20 cells, in general), favouring rapid mixing by numerical diffusion. We therefore need devoted numerical simulations of shock-cloud interactions (evolved versions of the one presented here) to study this process in detail. It would be of particular interest to study the dependence of cloud stability on the properties of the surrounding shocked gas, which shows very different thermodynamic properties for different jet powers (see, e.g., Figs.~\ref{fig:linesp} and \ref{fig:lines2}). These runs are, however, costly owing to the relatively small shock propagation velocity in clouds and out of the scope of this theoretical approach.




\subsection{Implications for large-scale feedback}
The presence of atomic outflows with large mass fluxes has been interpreted sometimes in the literature as jets being cold, strongly mass loaded and slow at large scales. This conclusion has been applied to study jet feedback at large scales \citep[beyond the host galaxy][]{2006ApJ...645...83V,2007MNRAS.376.1547C,2008MNRAS.384.1327S,2011MNRAS.415.1549G}, even for powerful outflows ($L_j \sim 10^{46}\,{\rm erg/s}$). In contrast, numerical simulations of powerful, relativistic jets show that it is very difficult to affect the global dynamics of powerful jets by stellar wind mass-load \citep{2014MNRAS.441.1488P,2021MNRAS.500.1512A}. In addition, observational studies have shown that such powerful jets must be mildly relativistic at large scales, on the basis of jet-to-counter-jet brightness ratios \citep[e.g.,][]{2004MNRAS.351..727A}.

Furthermore, the scenario we propose in this paper, which is based on results obtained from numerical simulations and is compatible with the physical scales of the problem, provides a reasonable frame that indicates that cold, massive outflows  observed in active galaxies are produced as a result of the development of relativistic jets, but do not represent the nature of jets as it was assumed by a number of works (see, e.g., the references above) for studying jet feedback at hundreds of kiloparsecs. The same conclusion has also been derived from observational results \citep[see, e.g.][who show that massive outflows can be explained from a small fraction of the total jet power or \citet{2002MNRAS.336.1161L} for the case of the FRI radio-galaxy 3C~31]{2014A&A...565A..46D}. Altogether, the possibility of massive, cold outflows with powers $\sim 10^{46}\,{\rm erg/s}$ not only not playing a role in large-scale feedback, but also not existing at all in the Universe, would have to be taken into serious consideration in cosmological or large-scale feedback simulations. In contrast, low power sources can be efficiently mass-loaded \citep{1996MNRAS.279..899B,2014MNRAS.441.1488P,2021MNRAS.500.1512A}, losing collimation and decelerating. However, the resulting outflows are not expected to reach hundreds of kiloparsecs as classical radio-galaxies do.

\section{Summary}

In this work, I present a general physical scenario to explain the presence of atomic gas outflows in GPS and CSS radio-sources. I have based the model in basic knowledge obtained from both numerical simulations and analytical models that have been systematically applied to analyse them \citep[radio-galaxy evolution, e.g.,][]{2007MNRAS.382..526P,2014MNRAS.445.1462P,2019MNRAS.482.3718P}.

The model predicts that atomic outflows could easily be detected in the GPS regime ($l_s \leq 1\,{\rm kpc}$) for low-power jets ($L_j \leq 10^{43}\,{\rm erg/s}$), as post-shock temperatures in clouds could easily fall below $10^4\,{\rm K}$. In addition, the cooling times of the shocked hot phase can be shorter than the evolution time-scales.

On the contrary, the post-shock temperatures reached in the case of powerful jets could ionise a large percentage of the existing atomic gas at these scales. Thus, the presence of atomic gas in powerful radio-galaxies at those scales requires rapid cooling in shocked clouds. This relies on cloud stability for sufficiently long periods of time, which seems to be granted by the small sound speeds expected for post-shock temperatures of $10^4-10^5\,$K (see Fig.~\ref{fig:lines1}). In this case, the cooling times allowing recombination to take place are compatible with the dynamical times required by jets to reach the CSS scales ($\geq 2\,{\rm kpc}$). 
In conclusion, a positive correlation between linear size and jet power should be expected in radio-galaxies showing atomic outflows.

Devoted numerical simulations of shock-cloud interaction must be run in order to study the stability and evolution of shocked cloud gas, also including the role of magnetic fields on the stability of the system \citep[see, e.g.,][for a recent work on the physics of shocked plasmas that could be relevant in this context]{2021A&A...645A..81S}. It has actually been shown that magnetic fields can be relevant to damp instability growth even if they are energetically irrelevant \citep{2022A&A...661A.117L}. If shocked clouds retain at least partly coherence, these regions can cool faster because of their larger densities and lower post-shock temperatures; in contrast, if instabilities and turbulent mixing with the shocked hotter ISM component, the resulting temperature of the mixture could delay cooling and put constraints to the validity of this simple model. Nevertheless, observational evidence seems to require shocked clouds to be fairly stable.

\bibliographystyle{aa.bst}
\bibliography{biblio}

\begin{acknowledgements} 
The author acknowledges support by the Spanish Ministry of Science through Grants PID2019-
105510GB-C31/AEI/10.13039/501100011033 and PID2022-136828NB-C43, from the Generalitat Valenciana through grant CIPROM/2022/49, and from the Astrophysics and High Energy Physics programme supported by MCIN and Generalitat Valenciana with funding from European Union NextGenerationEU (PRTR-C17.I1) through grant ASFAE/2022/005. The author also acknowledges the referee of this paper, Stas Shabala for his constructive criticism, which has improved the resulting text, and Pau Beltran-Palau for critical reading of the paper.
\end{acknowledgements}

\end{document}